\documentclass[12pt]{iopart}

\usepackage{iopams}  
\usepackage{color}  
\usepackage{graphicx}
\begin{document}

\title{Probing the neutrino mass ordering with \\ KM3NeT-ORCA: Analysis and perspectives}

\author{Francesco Capozzi$^{a}$, Eligio Lisi$^b$, Antonio Marrone$^{c,b}$}
\address{$^a$Department of Physics, Ohio State University,  Columbus, OH 43210, USA}
\address{$^b$Istituto Nazionale di Fisica Nucleare (INFN), Sezione di Bari, Italy}
\address{$^c$Dipartimento Interateneo di Fisica ``M.\ Merlin'', Universit\`a di Bari, Italy}
\ead{capozzi.12@osu.edu, eligio.lisi@ba.infn.it, antonio.marrone@ba.infn.it}
\vspace{10pt}

\begin{abstract}

The discrimination of the two possible options for the neutrino mass ordering (normal or inverted) is a
major goal for current and future  neutrino oscillation experiments. 
Such goal might be reached by observing high-statistics energy-angle spectra of events induced by
atmospheric neutrinos and antineutrinos propagating in the Earth matter. Large volume water-Cherenkov
detectors envisaged to this purpose include the so-called KM3NeT-ORCA project (in seawater) and 
the IceCube-PINGU project (in ice). Building upon a previous work focused on PINGU, we study in
detail the effects of various systematic uncertainties on the ORCA sensitivity  to the mass ordering,
for the reference configuration with 9~m vertical spacing.
We point out the need to control spectral shape uncertainties at the percent level, the
effects of better priors on the $\theta_{23}$ mixing parameter, and the benefits of 
an improved flavor identification in reconstructed ORCA events.
\end{abstract}


%
%
\submitto{\JPG\\ (Invited Contribution for Focus Issue on ``Neutrino mass and mass ordering'')}
%
%
%

\section{Neutrino mass ordering: Context}

The observation of neutrino flavor change in a variety of experimental settings has established that the three known neutrino families 
are characterized by three different mass states $\nu_i$ which mix with the flavor states $\nu_\alpha$ \cite{PDG}. 
In particular, two squared mass difference ($\delta m^2,\,\Delta m^2)$ and three mixing angles ($\theta_{12},\,\theta_{13},\,\theta_{23}$) have been
measured \cite{Ca17}. The unknown sign of $\Delta m^2$ is physical and distinguishes the cases of so-called ``normal ordering'' (NO, positive $\Delta m^2$) and ``inverted ordering'' (IO, negative $\Delta m^2$) for the neutrino mass spectrum $(m_1,\,m_2,\,m_3)$.%
\footnote{We use the notation $\delta m^2=m^2_2-m^2_1>0$ and $\Delta m^2=m^2_3-(m^2_2+m^2_1)/2$ as in
\protect\cite{Ca17,Fo05}.}

The $\nu$ mass ordering can be experimentally probed in various ways. In flavor oscillation experiments, one seeks the interference 
of $\Delta m^2$-driven oscillations with some $Q$-driven oscillations, where $Q$ is another quantity (with squared-mass dimensions) having a known sign,
say, $Q>0$. 
The interference pattern between $Q$ and $\Delta m^2$ can then reveal the sign of the latter. At present, one may envisage three possibile $Q$'s 
and thus three observational opportunities: (1) $Q=\delta m^2$, testable with medium-baseline reactor $\nu$ oscillations in vacuum \cite{Reac}; 
(2) $Q=V_e$, where $V_e=\sqrt{2}G_F N_e$ is the effective potential on background electrons with density $N_e$, testable within the Earth matter 
by atmospheric and long-baseline accelerator $\nu$ experiments \cite{Matt}; and (3)
$Q=V_\nu$, where $V_\nu=\sqrt{2}G_F N_\nu$ is the effective potential on background neutrinos (or antineutrinos) with density $N_\nu$, 
testable during the core-collapse event of a galactic supernova \cite{SNova}. 

Each of above three opportunities faces different challenges \cite{Voge}. 
In case (1), the energy resolution must reach the level  $\delta m^2/|\Delta m^2|\simeq 3\%$, and the energy scale must be controlled at the 
subpercent level, in order to probe with sufficient accuracy the amplitude and the phase of the interference term, respectively. In case (2), 
potentially large mass-ordering features are suppressed in observable atmospheric $\nu$ spectra by the finite energy-angle resolution and by 
the combination of appearance-disappearance and $\nu$-$\overline\nu$ channels, while for accelerator $\nu$ such features are entangled with other oscillation effects,
most notably those induced by a possible CP-violation phase $\delta$.  Finally, the rare event associated to case (3) represents both an  
experimental dream and a theoretical challenge, since the physics of $\nu$ oscillations in a dense neutrino gas 
seems exceedingly difficult to be captured in realistic situations. In all cases, very high statistics will be needed to see 
mass-ordering features emerge in event spectra. 

Nonoscillation experiments probe instead absolute $\nu$ masses, either kinematically or dynamically. Also in this context there are three opportunities
\cite{Abso}:
(1) Beta-decay endpoint tests, which represent
a classical kinematic method, sensitive to the effective parameter $m^2_\beta=\sum_i |U_{ei}|^2 m^2_i$; (2) searches for neutrinoless double beta decay 
$0\nu\beta\beta$ (only for Majorana neutrinos), sensitive to the effective parameter $m_{\beta\beta}= |\sum_i U^2_{ei} m_i|$; and (3) precision
cosmology, sensitive (in first approximation) to the total ``gravitational charge'' of neutrinos, namely, $\Sigma=\sum_i m_i$. Unless the 
mass states are quasi-degenerate ($m_i \gg \sqrt \Delta m^2$, disfavored but not yet excluded by current data) these absolute mass observables
may provide additional handles to constrain the mass ordering. 

At present, a global analysis of the available oscillation and nonoscillation data suggest a slight overall preference for NO over IO, at the level of 
$\sim 2\sigma$, mainly driven by Super-Kamiokande atmospheric neutrino data \cite{Ca17}. This hint, although too weak to draw any conclusion,
illustrates the physics potential of atmospheric neutrino experiments in probing the mass ordering via matter effects at $O(1\div10)$~GeV energies. 
For this class of experiments, reaching a mass-ordering sensitivity $\geq\! 3\sigma$ requires new-generation detectors with both large volume (to increase statistics) and dense instrumentation (to probe relatively low energies with good angular resolution). 
In this context, two dedicated projects have recently been put forward. On the one hand, the Precision IceCube Next Generation Upgrade (PINGU) has been proposed as a low-energy in-fill extension to the IceCube Observatory at the South Pole, based on the Cherenkov detection technique in ice \cite{PINGU}. 
On the other hand, the KM3NeT collaboration has proposed the Oscillation Research with Cosmics in 
the Abyss (ORCA), a new low-energy neutrino telescope based on the Cherenkov detection technique in seawater, offshore from Toulon, France 
\cite{ORCA}.
Each of the two projects is expected to achieve a NO-IO discrimination $\geq 3\sigma$ in a few years of data taking \cite{PINGU,ORCA}. 

As in Super-Kamiokande, the main observables in PINGU and ORCA are represented by the double differential distributions of neutrino events in energy and 
zenith angle, averaged over the azimuthal angle. 
In principle, the energy-angle distributions of electron and muon (anti)neutrinos at the detector site 
would encode major oscillation effects both in vacuum (above the horizon) and in matter (below the horizon)
\cite{Oscil,Malt}. However, one can only observe 
(part of) the final state induced by $\nu$ interactions in the effective volume of the detector, and a significant degradation of
mass-ordering effects in matter occurs when passing from unobservable neutrino spectra to observable event spectra. In particular:
(a) neutrino and antineutrino events cannot be distinguished; (b) electron and muon flavors are only probed via proxies, e.g.\ via the categories 
of ``cascade'' (or ``shower'') and ``track'' events, respectively, 
with inherent flavor mis-identification and contamination issues; (c) the reconstructed values of the  
energy $E$ and zenith angle $\theta$ of the parent neutrino are largely smeared around the (unobservable) true values, 
due to unavoidable $\nu$ scattering and detection effects. Further smearing arises from uncertainties in the 
neutrino source (atmospheric fluxes), propagation (oscillation parameters) and detection (cross sections, effective volumes)
As a result, the difference between NO and IO in observable event spectra is reduced to mild 
spectral variations, which typically amount to a few percent and require to be summed
over many bins to become apparent in a statistical analysis (see e.g.\ \cite{PINGU,ORCA,Oscil}). In other words,
mass-ordering features will not appear at a glance in the measured spectra, 
but will emerge from careful statistical analyses, based on 
a detailed comparison of the NO and IO spectral templates with the real data: small ($\ll 1\sigma$) differences 
in many individual bins will eventually sum up to $>3\sigma$ differences in the whole spectrum.  

In this context, it makes sense to investigate the effects of systematic uncertainties of observable spectra on the (minimum)
experimental sensitivity to the mass ordering. Indeed, for a given mass-ordering sensitivity estimate (say, ``$n\sigma$ in $m$ years''), 
the addition of any single systematic error or nuisance parameter can only {\em decrease\/} such estimate, even if just by a tiny amount.
Including  many possible and diverse sources of uncertainties in a dedicated analysis
can thus be helpful to identify major systematic effects on the mass ordering sensitivity. In this paper we perform such analysis 
for the KM3NeT-ORCA project, building upon a previous work which focused on PINGU \cite{Build}. 
In our opinion, the differences between the two projects in various aspects 
(including energy and angle resolution, effective volume, flavor identification) and the
increasing interest in the neutrino community on next-generation atmospheric detectors warrant a new detailed study, focused on the ORCA official proposal as
presented in \cite{ORCA}.

The literature on ORCA (apart from the KM3NeT collaboration studies) is relatively limited as compared to PINGU. A study of the ORCA (and PINGU) atmospheric neutrino sensitivity to mass ordering was performed in \cite{Winter}, prior to the proposal in \cite{ORCA} and with less systematic error sources as compared to this work. An alternative approach via accelerator neutrinos beamed to ORCA was considered in \cite{Razzaque}. For previous PINGU studies the reader is referred to the bibliography in \cite{Build}, including \cite{Ble1,Ble2,Ge}. A preliminary comparison of PINGU and ORCA in terms of detector characteristics and expected performances was discussed in \cite{Janez}.

This work 
	focuses on the ORCA project as presented in \cite{ORCA}, and it 
is structured as follows. In Section~2 we describe the input and methodology adopted in our analysis of the
ORCA sensitivity to mass ordering.
In Section~3 we present detailed results and discuss the most important sources of systematic uncertainties. 
We also argue that improvements in flavor identification would be most beneficial to ORCA. 
We summarize our findings in Section~4.

\section{Prospective ORCA Analysis: Methodology and Inputs} 

The main observables in ORCA are the double differential event spectra in energy $E$ and zenith angle $\theta$. The calculation of such distributions
is performed with the same methodology adopted for PINGU in \cite{Build}, to which we refer the reader for notation and details not repeated herein. We just highlight the main differences with respect to \cite{Build}. 
 
Concerning the neutrino source, we take atmospheric neutrino fluxes as calculated in \cite{Fluxes} for the Frejus site (without mountain). 
Concerning neutrino propagation, we assume as reference (true) oscillation parameters the following set, slightly updated from 
\cite{Build} 
as a result of a recent global data analysis \cite{Ca17} 
(see also \cite{Concha,Valle}):
\begin{eqnarray}
\label{Dm2} |\Delta m^2|_{\mathrm{true}} &=& 2.5\times 10^{-3}\ \mathrm{eV}^2\ , \\
\label{dm2} \delta m^2|_{\mathrm{true}} &=& 7.5\times 10^{-5}\ \mathrm{eV}^2\ , \\
\label{q13} \sin^2\theta_{13}|_{\mathrm{true}} &=& 0.0215 \ ,\\
\label{q12} \sin^2\theta_{12}|_{\mathrm{true}} &=& 0.308 \ ,\\
\label{dcp} \delta|_{\mathrm{true}} &=& 3\pi/2 \ .
\end{eqnarray}
For the uncertain parameter $\theta_{23}$, unless stated otherwise, we assume as in 
\cite{Build} that it can take any true value in the range 
\begin{equation}
\label{q23} \sin^2\theta_{23}|_{\mathrm{true}} \in [0.4,\,0.6]\ .
\end{equation}

Concerning detection, the main ORCA characteristics are taken from 
\cite{ORCA} in digitized form, including energy and angular resolutions and effective volumes 
for different classes of events as a function of the parent neutrino energy, for the benchmark configuration with $9$~m vertical spacing.%
\footnote{
We thank A.\ Kouchner and J.\ Brunner for
useful discussions and for providing us with numerical parametrizations of the ORCA detector characteristics
graphically shown in \protect\cite{ORCA}. We understand that assessing more precisely such characteristics represents 
work in progress, and that numerical parametrizations should be considered as intermediate results subject to updates.}

\begin{figure}[t]
\hspace*{2.2cm}
\includegraphics[height=6.9cm]{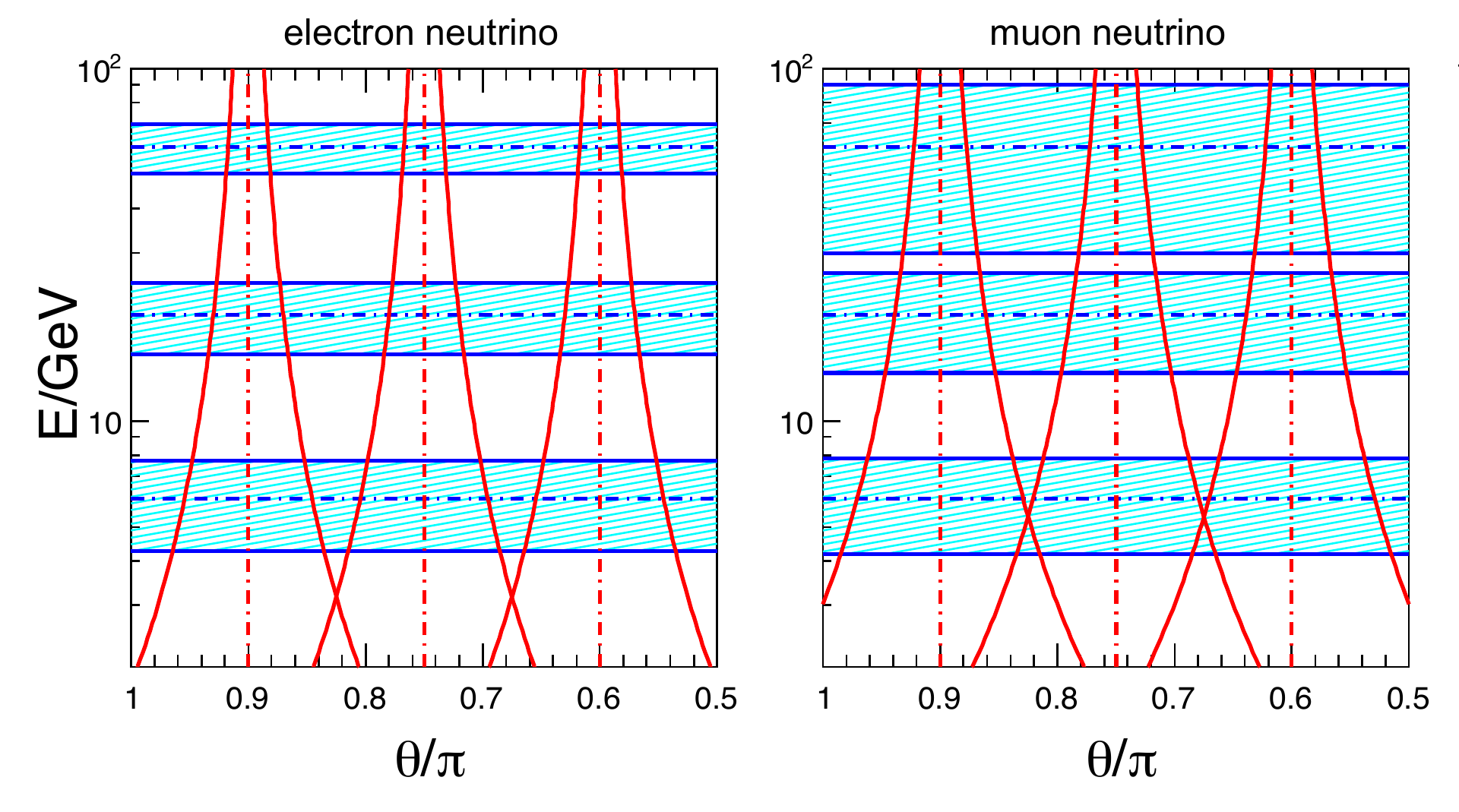}
\vspace*{-4mm}
\caption{Widths (at $\pm1\sigma$) of the resolution functions in energy and angle for events induced 
by $\nu_e$ (left) and by $\nu_\mu$ (right), in terms of logarithmic energy versus the zenith angle, for 
three representative values in both coordinates. Resolutions for antineutrino events (not shown) are similar.   
\label{Fig_01}}
\end{figure}

Typical angular and energy resolution widths (at $\pm 1\sigma)$ for ORCA are shown in Fig.~\ref{Fig_01}, to be compared with the ones for PINGU in
\cite{Build}. The plane in Fig.~\ref{Fig_01} is charted  by the true neutrino energy (in logarithmic scale) and by the true neutrino zenith angle
$\theta$ (in linear scale), where $\theta/\pi=1$ and 0.5 correspond to upgoing vertical and horizontal directions, respectively.%
\footnote{We prefer to use the variable $\theta$ rather than $\cos\theta$, as motivated in \cite{Build}.} 
The left and right panels refer to electron and muon neutrino events in ORCA.
 The hatched horizontal bands correspond to the ($\theta$-independent) energy resolution widths $\pm\sigma_E$ for the representative values
$E= 6$, 20 and 60~GeV. The vertical curves correspond to the ($E$-dependent) angular resolution widths $\pm\sigma_\theta$ for the representative values
$\theta/\pi=0.6$, 0.75 and 0.9. As compared with PINGU \cite{Build,Janez}, the angular resolution is generally expected to be
improved in ORCA, while the energy resolution may or may not be improved, depending on the neutrino flavor and on energy range. 
In particular, an improvement is expected in ORCA for the energy resolution of $\nu_e$ but not of $\nu_\mu$ events (especially at high energy).
However, it is difficult to gauge a priori the impact of such differences between PINGU and ORCA, and a thorough investigation is needed.  
In the analysis, we take $E \in [2,\,100]$~GeV and $\theta/\pi\in[0.5,\,1]$ as in Fig.~\ref{Fig_01}, each of these two ranges being
divided into 20 equally-spaced bins (linearly for $\theta$ and logarithmically for $E$), for a total of $400+400$ bins of cascade~+~track distributions.

A relevant difference with respect to \cite{Build} is that we no longer identify ``cascade'' and ``track'' events with ``$\nu_e+\overline\nu_e$'' and ``$\nu_\mu+\overline\nu_\mu$'' flavor events, respectively. The ORCA proposal \cite{ORCA} includes a simulation of the probability that cascade and track event samples  contain ``wrong flavor'' events, as well as $\nu_\tau$ and neutral current events, as a function of energy. Using ORCA inputs \cite{ORCA} we compute and include such event contaminations in the analysis of both cascade- and track-event distributions. We shall also comment on the effect of ideally removing any such contamination.  

\begin{figure}[t]
\hspace*{2.2cm}
\includegraphics[height=6.0cm]{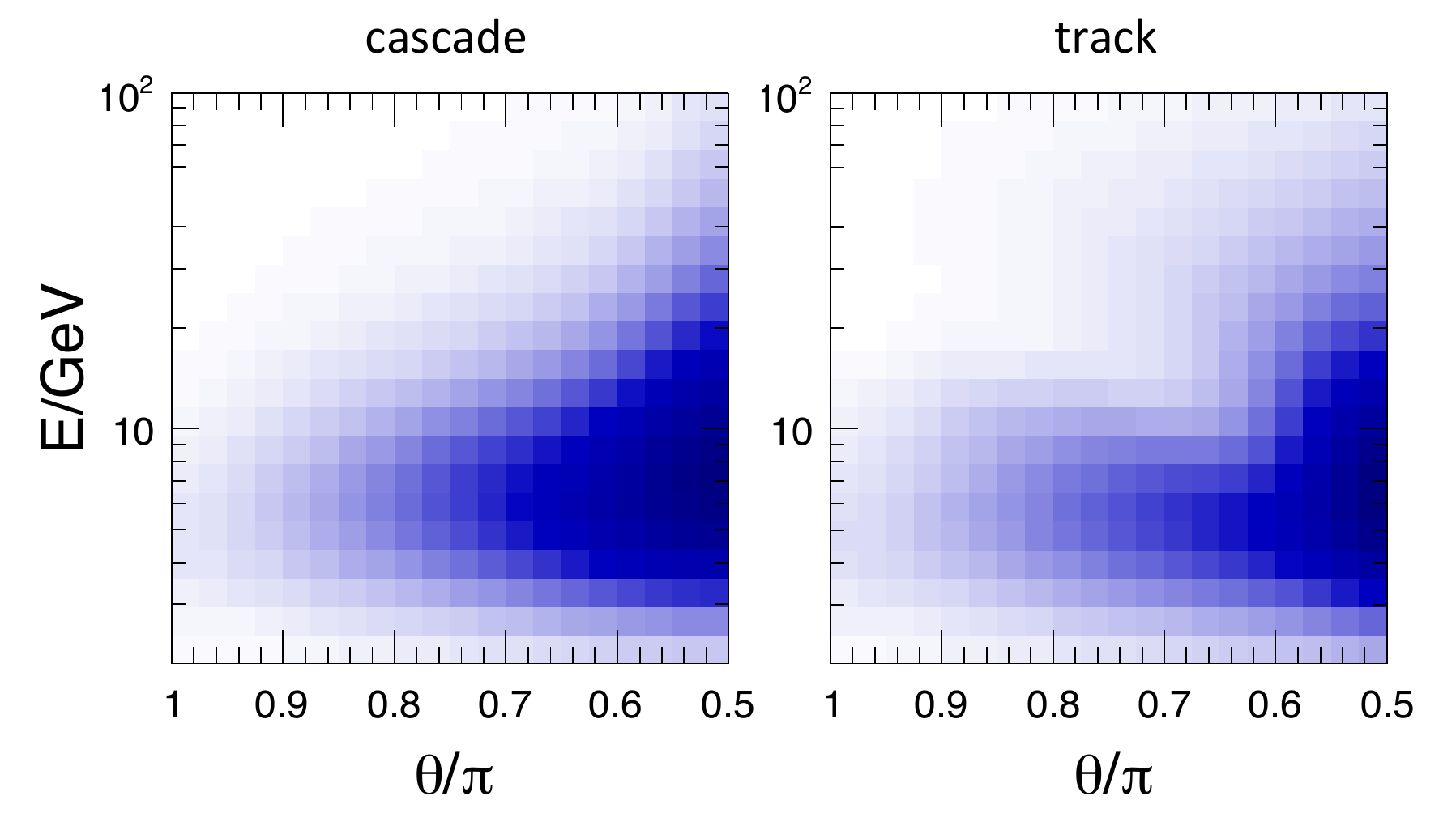}
\vspace*{-4mm}
\caption{Energy-angle distributions of cascade (left) and track (right) events in ORCA, for reference oscillation parameters in NO and 
$\sin^2\theta_{23}=0.5$. Color intensity is proportional to the number of events in each bin (in arbitrary units). 
The corresponding distributions in IO (not shown) would be indistinguishable  by eye.   
\label{Fig_02}}
\end{figure}

Figure~\ref{Fig_02} shows an example of our calculated distributions of cascade and track events in ORCA, for the above reference oscillation parameters
and $\sin^2\theta_{23}=0.5$ in NO. As already commented in \cite{Build} for PINGU, the event distributions are peaked towards the horizon,
where atmospheric neutrino fluxes become larger but, unfortunately, matter effects become smaller. Moreover, the
distributions are quite featureless, the only visible oscillation remnant being a slanted ``valley'' in the track event spectrum, 
induced by the first oscillation minimum \cite{Build}. Mass-ordering signatures are largely smeared out, and
the corresponding distributions for IO (not shown) would be visually indistinguishable from those in Figure~\ref{Fig_02}. 
Spectral NO-IO differences remain typically in the few percent range and do not emerge ``by eye'': they 
must be extracted by a statistical analysis of the data, including detailed estimates of systematic uncertainties.

The statistical analysis of ORCA event distributions is performed through the $\chi^2$ approach described in \cite{Build}, leading to
a ``number of sigma'' estimator $N_\sigma=\sqrt{\Delta \chi^2}$, which quantifies the separation between the true ordering (TO, data)
and wrong ordering (WO, test), where either T=N and W=I, or T=I and W=N. We build the $\chi^2$ function via an increasingly rich set of 
systematic uncertainties, in addition to the statistical ones: 
\begin{itemize}
\item  oscillation and normalization uncertainties (minimal set),
\item  plus energy-scale and energy-angle resolution uncertainties (representing ``known'' spectral systematics),
\item plus energy-angle spectral shape uncertainties (via quartic polynomials,  representing ``poorly known'' spectral systematics),
\item plus residual uncorrelated systematics in each bin (maximal set of uncertainties).
\end{itemize}

Oscillation uncertainties include the fractional errors of $\Delta m^2$ and $\theta_{13}$, that we update from \cite{Build} following 
\cite{Ca17}:
$\sigma(\Delta m^2)=1.6\%$ and $\sigma(\sin^2\theta_{13})=  4.0\%$.
 The parameter $\delta$ is fixed at $3\pi/2$ for TO, while it is 
left free in the range $[0,\,2\pi]$ for the WO. 
The parameter $\sin^2\theta_{23}$ is taken in the range $[0.4,\,0.6]$ for TO, while
it is left free in the whole range $[0,\,1]$ for WO.
The parameters $\delta m^2$ and $\sin^2\theta_{12}$ are kept fixed, their uncertainties being 
negligible in the ORCA analysis. 

Normalization uncertainties include, as in \cite{Build}, an overall normalization error ($15\%$), 
the relative $\mu/e$ and $\overline\nu/\nu$ flux uncertainties ($8\%$ and $5\%$, respectively),
and the Earth's core density error ($3\%$). 

The fractional uncertainties affecting the resolution functions are also estimated as in \cite{Build}. We assume that the energy resolution function 
may be biased (up to $5\%$ at $1\sigma$), and that the energy and angle resolution widths may be affected by a $10\%$  errors, independently 
for cascade and track events. 

As in \cite{Build}, we provide allowance for further, generic 
shape variations of the energy-angle spectra of cascade and track events, with a typical size of $\pm 1.5\%$;
we also consider the cases of halved (0.75\%) and doubled ($3\%$) shape errors. These errors are meant 
to characterize systematic effects which are known to exist, but whose energy or angular dependence is poorly or incompletely known,
including: uncertainties in the primary cosmic ray fluxes, differential atmospheric neutrino fluxes and cross sections and,
to some extent, energy-angle detection efficiencies. Spectral uncertainties are 
parametrized via polynomial deformations with constrained coefficients (the zeroth order coefficient being removed, to avoid
double counting of the overall normalization error). While in \cite{Build} we considered polynomials from linear (minimal case,
corresponding to tilted spectra) to quartic terms, for the sake of simplicity we consider only the full case with quartics, which
leads to more conservative results.

 We emphasize that, since the work \cite{Build}, the case for conservative estimates of spectral
uncertainties has been growing. Atmospheric neutrino flux uncertainties in energy and angle have been largely discussed 
in a dedicated workshop \cite{Atm1}, following earlier suggestions in \cite{Atm0}. Building upon \cite{Barr},
recent developments have been presented in \cite{Fedy}, which may eventually lead to covariance matrices for atmospheric 
neutrino flux errors from different systematic sources, at least in the energy domain. 
The envelope of the error estimates in \cite{Fedy} can be sizable in the energy range 
relevant for ORCA, and its functional dependence cannot be captured by a simple first-order expansion in energy (i.e., by a spectral tilt).
At the same time, it is being increasingly recognized that total and differential $\nu$ cross sections are not known
at the level required for precision physics \cite{Cross}, and that their current uncertainties should not be underestimated in spectral analyses.
We thus surmise that the conservative approach to 
spectral shape uncertainties proposed in \cite{Build} for PINGU is still justified and can be usefully applied also to ORCA.

Finally, we consider the possibility of residual,  bin-to-bin uncorrelated errors. A classical example is
provided by the finite Monte Carlo statistics of experimental simulations. In our case the test
spectra are calculated, but in a real experiments they are necessarily simulated. 
As in \cite{Build}, we assume additional $1.5\%$ errors in each bin (uncorrelated),
but consider also the case of halved ($0.75\%$) and doubled ($3\%$) size for such systematics.

\section{Prospective ORCA Analysis: Results and Discussion}

\begin{figure}[t]
\hspace*{2.4cm}
\includegraphics[height=7.2cm]{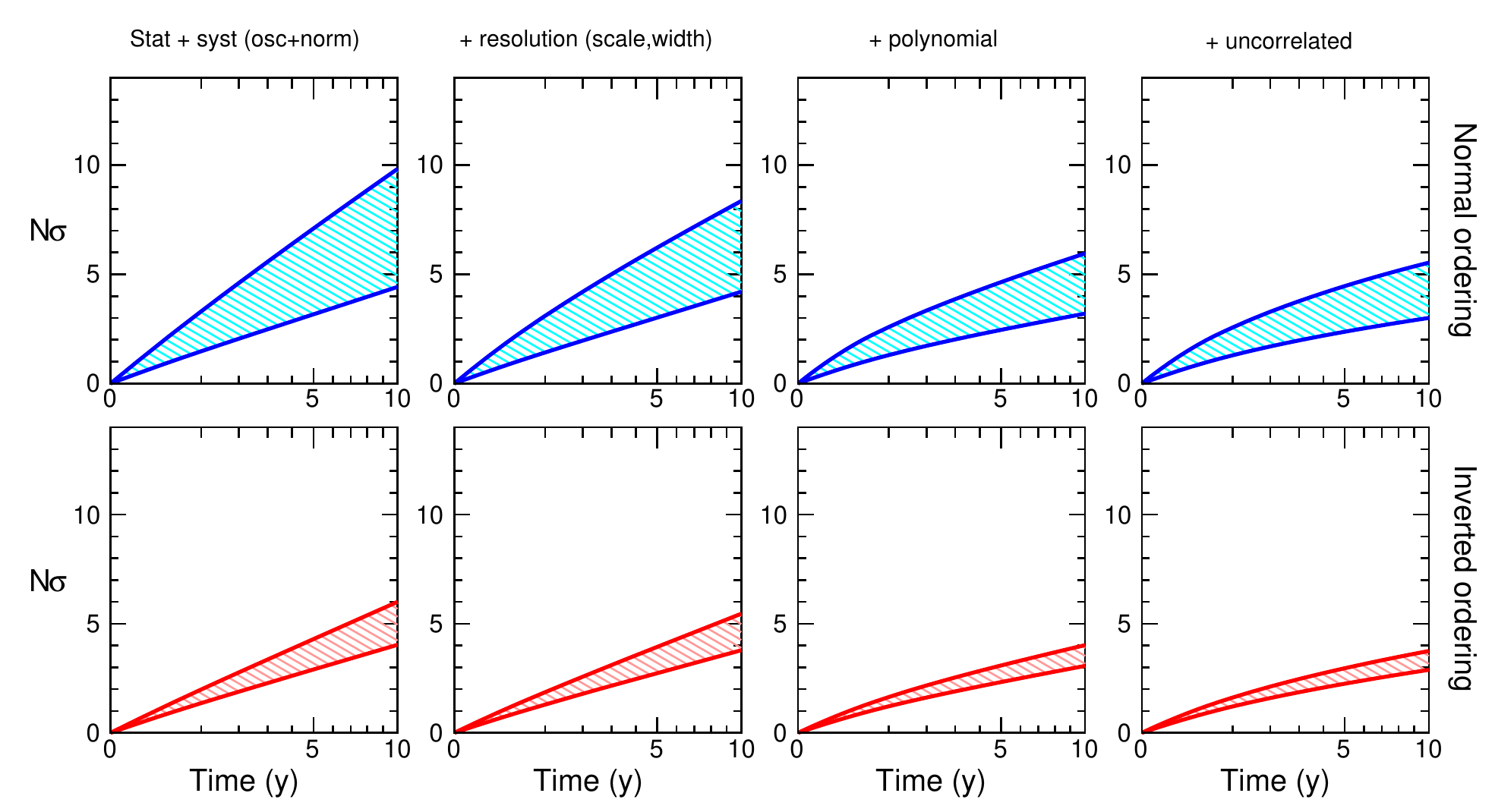}
\vspace*{-7mm}
\caption{Estimated ORCA sensitivity to mass ordering ($N_\sigma$), 
for either true NO (top panels) or true IO (bottom panels), in terms of live time $T$ (years). 
From left to right, the fit includes the following
systematic errors: oscillation and normalization uncertainties, energy scale and resolution width errors, polynomial shape systematics
(with quartic terms) at the 1.5\% level, and uncorrelated systematics at the 1.5\% level. 
Note that the abscissa scales as $\sqrt{T}$ to 
emphasize the effect of systematics.
Bands corresponds to the envelope of all
sensitivity lines for  $\sin^2\theta_{23}|_\mathrm{true} \in[0.4,\,0.6]$. 
\label{Fig_03}}
\vspace*{-0mm}
\end{figure}

Figure~\ref{Fig_03} shows the ORCA sensitivity to the mass ordering, in terms of 
$N_\sigma$ separating the true ordering (top: NO; bottom: IO)
from the wrong ordering, as a function of
the detector live time $T$ in years. The sensitivity would be represented by a line for fixed $\sin^2\theta_{23}|_\mathrm{true}$; by spanning the 
range $\sin^2\theta_{23}|_\mathrm{true} \in[0.4,\,0.6]$, sensitivity bands are obtained as envelopes. 
The abscissa is scaled as $\sqrt{T}$, so that the sensitivity would grow linearly in the ideal case of statistical
errors only. From left to right,  the fit includes an increasingly rich error set, as detailed in the previous section: 
oscillation and normalization errors, energy scale and resolution width errors, spectral shape uncertainties
(via quartic polynomial), and uncorrelated systematics. The last two errors are kept at the default level of 1.5\%. 
With just oscillation and normalization errors, the $N_\sigma$ bands grow almost linearly in $\sqrt{T}$, even after 10 years of data taking. To some 
extent, this behavior persists also by including resolution uncertainties, with only a minor decrease of sensitivity, implying that ORCA 
is not limited by such systematics. However, when spectral shape variations are allowed to float at the $1.5\%$ level 
(via quartic polynomial parametrization), the bands are bent and the sensitivity is noticeably degraded. The further    
addition of uncorrelated systematics at the same level ($1.5\%$) worsens the sensitivity only slightly. 

From Fig.~\ref{Fig_03} 
we learn that the ORCA sensitivity may be degraded by possible spectral deformations 
 from poorly understood or controlled systematic sources 
(related to atmospheric fluxes, cross sections, or instrumental effects), already at the 1.5\% level of fractional variations. 
The possibility of such spectral variations should not be overlooked, as demonstrated 
by the emergence of unexpected (and still unexplained)  features in the context of reactor $\nu$ spectra \cite{Bump},
which were supposedly well-known at the $\sim\! 2\%$ level.  

\begin{figure}[t]
\hspace*{2.4cm}
\includegraphics[height=7.2cm]{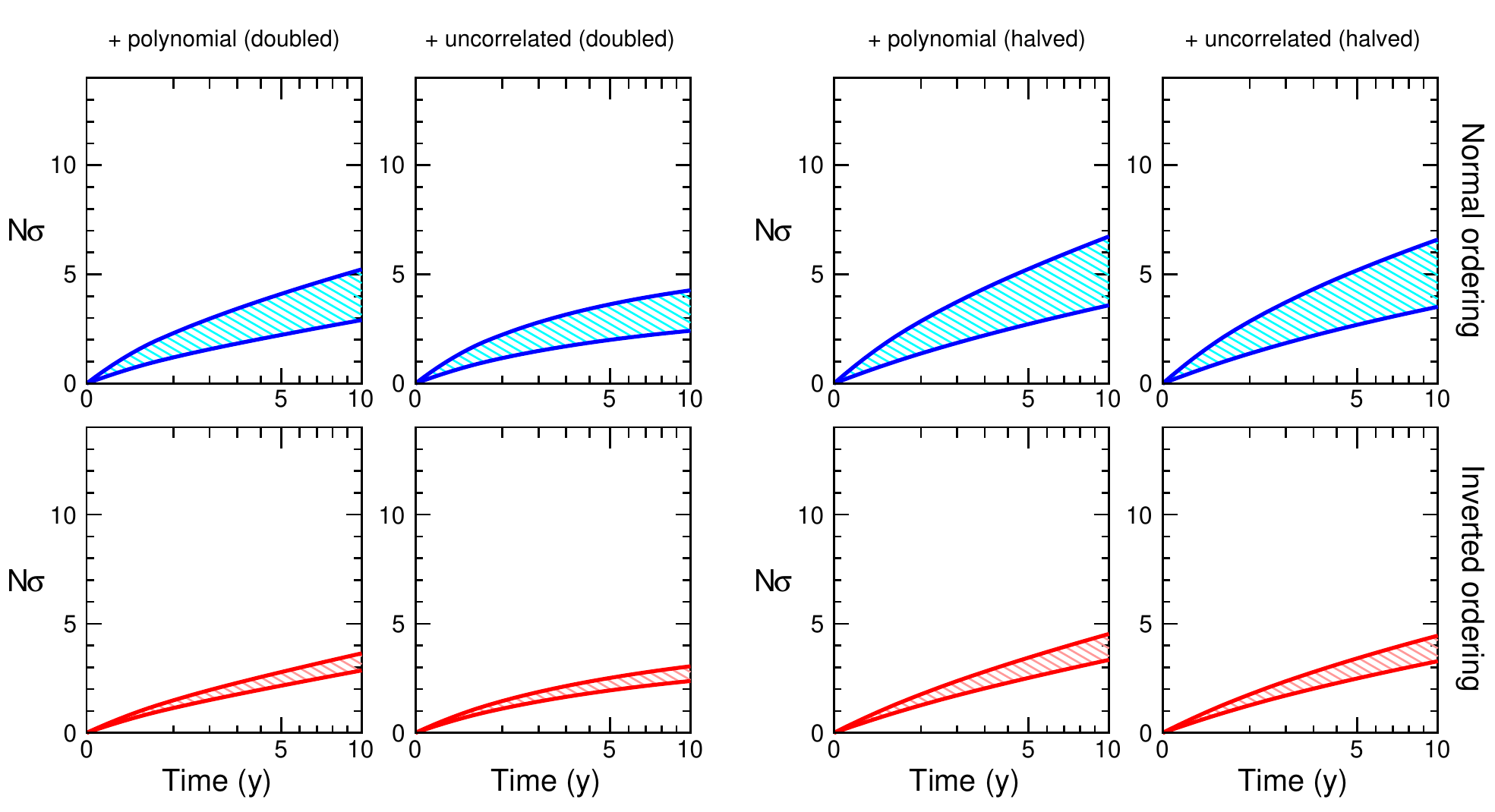}
\vspace*{-7mm}
\caption{As in the last two panels of Fig.~\protect\ref{Fig_03}, but for doubled size ($3\%$, left)
and halved size ($0.75\%$, right) of correlated polynomial and uncorrelated systematics.  
Statistical
errors and systematics related to oscillation, normalization and resolution uncertainties are assumed to be the same. 
\label{Fig_04}}
\end{figure}

Figure~\ref{Fig_04} shows the results of our analysis as obtained by 
either doubling (left) or halving (right) the size of polynomial and uncorrelated systematics, 
as compared to the default level of $1.5\%$ in Fig.~\ref{Fig_03}. It appears that allowance for few-percent spectral  variations
would  induce systematic limitations to the experiment.
Thus, it is important to reach a control of the spectral systematics at the (sub)percent level.

\begin{figure}[t]
\hspace*{2.4cm}
\includegraphics[height=7.2cm]{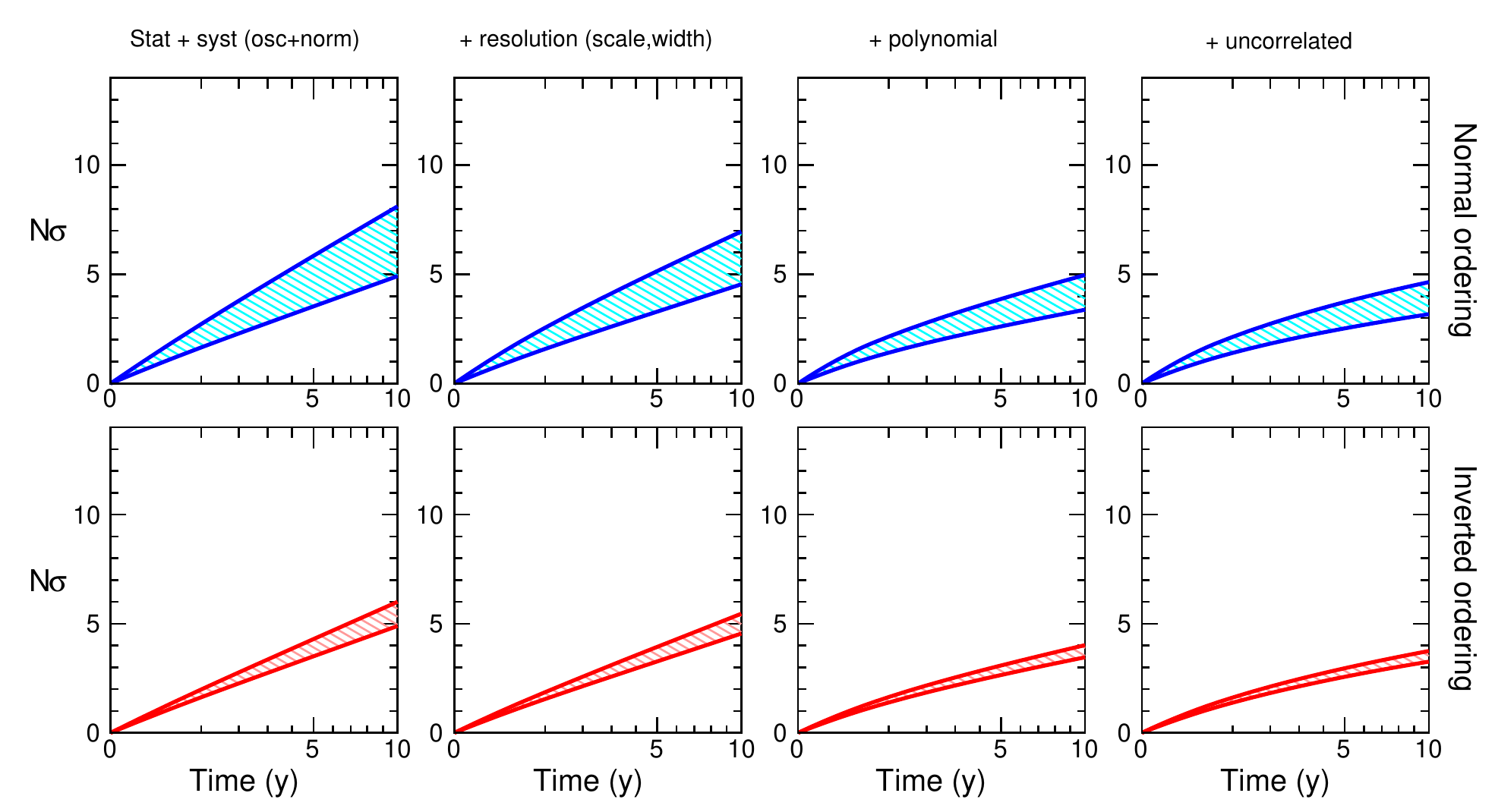}
\vspace*{-7mm}
\caption{As in Fig.~\protect\ref{Fig_03}, but for the restricted range $\sin^2\theta_{23}|_\mathrm{true} \in[0.46,\,0.54]$.   
\label{Fig_05}}
\end{figure}

Besides mitigating the effect of spectral shape uncertainties, another possible direction for improving the ORCA sensitivity 
to mass ordering is the reduction of oscillation parameter uncertainties, most notably of $\theta_{23}$. 
Figure~\ref{Fig_05} shows the sensitivity bands for $\theta_{23}$ spanning the reduced range 
$\sin^2\theta_{23}|_\mathrm{true} \in[0.46,\,0.54]$. There is a noticeable gain with respect to the corresponding sensitivities
in Fig.~\ref{Fig_03}. It can be expected that the currently large uncertainties in $\theta_{23}$ \cite{PDG}, partly
related to the difficulty of lifting the octant degeneracy \cite{Ca17}, will be reduced by new data in the next future \cite{Accel}. 
Of course, ORCA will also provide its own measurement of $\theta_{23}$, besides testing the mass ordering. However,
the two observations are somewhat entangled, and the uncertainty in one of them affects the other, as shown below.

\begin{figure}[b]
\hspace*{2.4cm}
\includegraphics[height=9.2cm]{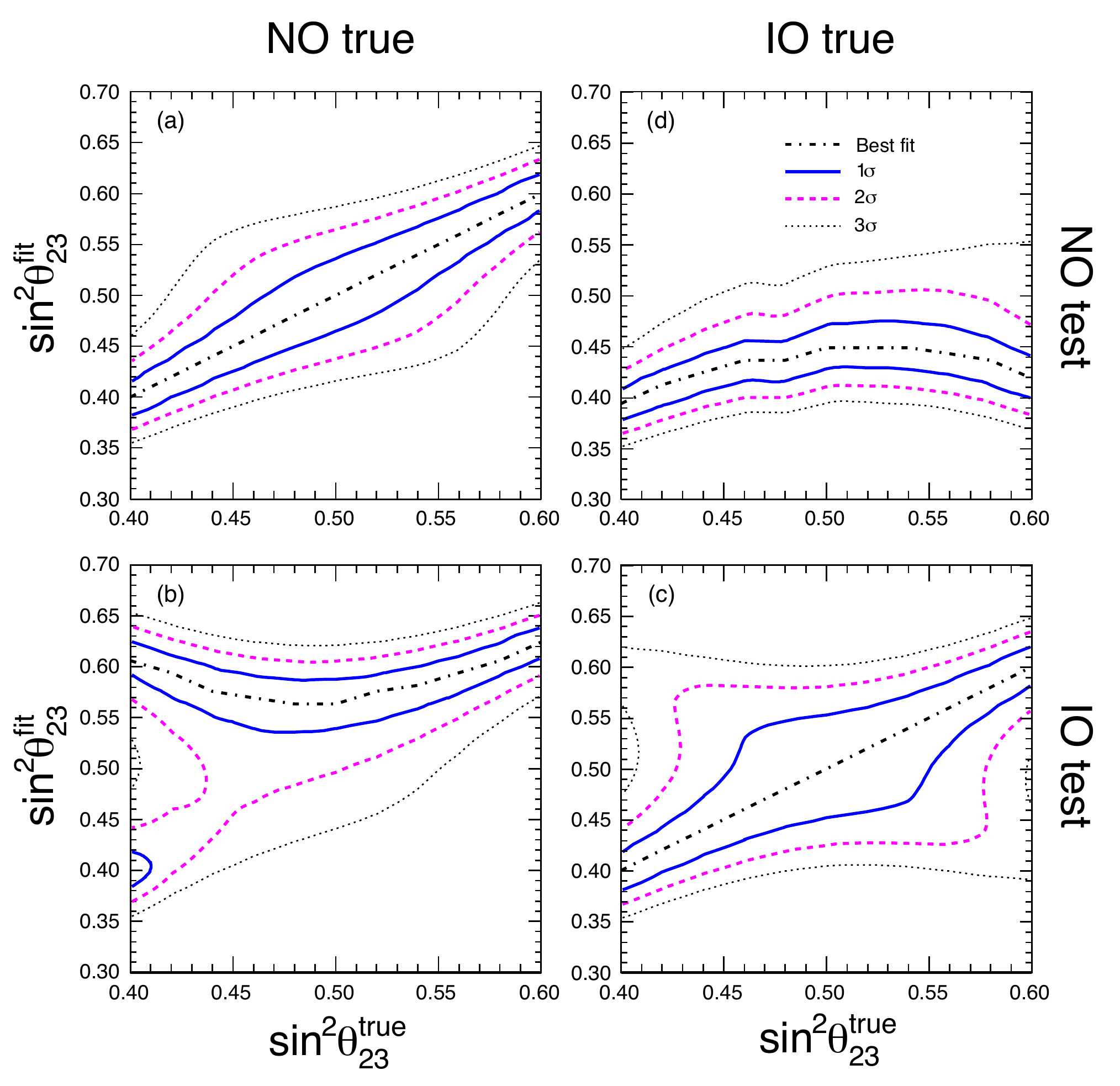}
\vspace*{-1mm}
\caption{Fitted value $\sin^2\theta^\mathrm{fit}_{23}$ (at 1, 2 and $3\sigma$) versus the true value
$\sin^2\theta^\mathrm{true}_{23}$, for the four possible cases where the test ordering (i.e., the one assumed in the
fit) is either the true or the wrong one:
(a) NO~=~true, NO~=~test; (b) NO~=~true, IO~=~test; (c) IO~=~true, IO~=~test; (d) IO~=~true, NO~=~test. 
\label{Fig_06}}
\end{figure}
Fig.~\ref{Fig_06} shows, in each panel, the fitted value $\sin^2\theta^\mathrm{fit}_{23}$ (at 1, 2 and $3\sigma$) as a function of the true value
$\sin^2\theta^\mathrm{true}_{23} \in [0.4,\,0.6]$, for the four possible cases where the true and tested ordering coincide or not in ORCA. 
The results refer to default systematic errors and to a 5-year exposure. The ``diagonal'' panels (with identical mass ordering for the
true and test cases) show that the measured $\theta_{23}$ is highly correlated with the true one, especially for NO, while
for IO the octant ambiguity affects the measurements at 2--3$\sigma$ level.  The ``off-diagonal'' panels (with different ordering
for true and test cases) show instead that the measured $\theta_{23}$ is largely decoupled (and generally different) from the true value.
As far as the true mass ordering is not determined, measuring $\theta_{23}$ in ORCA will thus remain affected by large uncertainties and by 
possible biases. 

The above results clearly show that the ORCA sensitivity to mass ordering will benefit
from any improvement in constraining systematic shape errors and oscillation parameter uncertainties 
(in particular of $\theta_{23}$). We conclude this section by discussing other ways to improve the sensitivity.

One possibility is to increase the effective volume via smart reconstruction strategies and loosened triggers, which could lead to
a gain factor $\sim\!1.5$ in exposure, as recently discussed in \cite{Kouchner,Coelho}. According to preliminary investigations,
this gain is expected to be quite uniform in energy and to leave unaltered other detection characteristics, such as the energy 
threshold and the resolution widths \cite{Kouchner,Coelho}.  In such conditions our results in Figs.~3--5 would still hold, but with 
a time scale reduced by an overall factor  $\sim\!0.67$.  

Another relevant improvement would emerge from mitigating flavor mis-identification and contamination effects. As already mentioned,
cascade and track events do not necessarily coincide with events generated by charged-current (CC)
interactions of electron and muon neutrinos, respectively, due to possible cross-talk between these two samples and to backgrounds from
tau-flavor and NC events. The mis-identification and contamination fractions, which may be substantial, have been taken into account in
all the above results, following the ORCA event classification energy profiles in \cite{ORCA}. 
Note that flavor cross-talk effects tend to dilute the mass-ordering signatures, 
which are mainly contained in the muon-to electron neutrino appearance channel \cite{Matt}. 

\begin{figure}[t]
\hspace*{2.4cm}
\includegraphics[height=7.2cm]{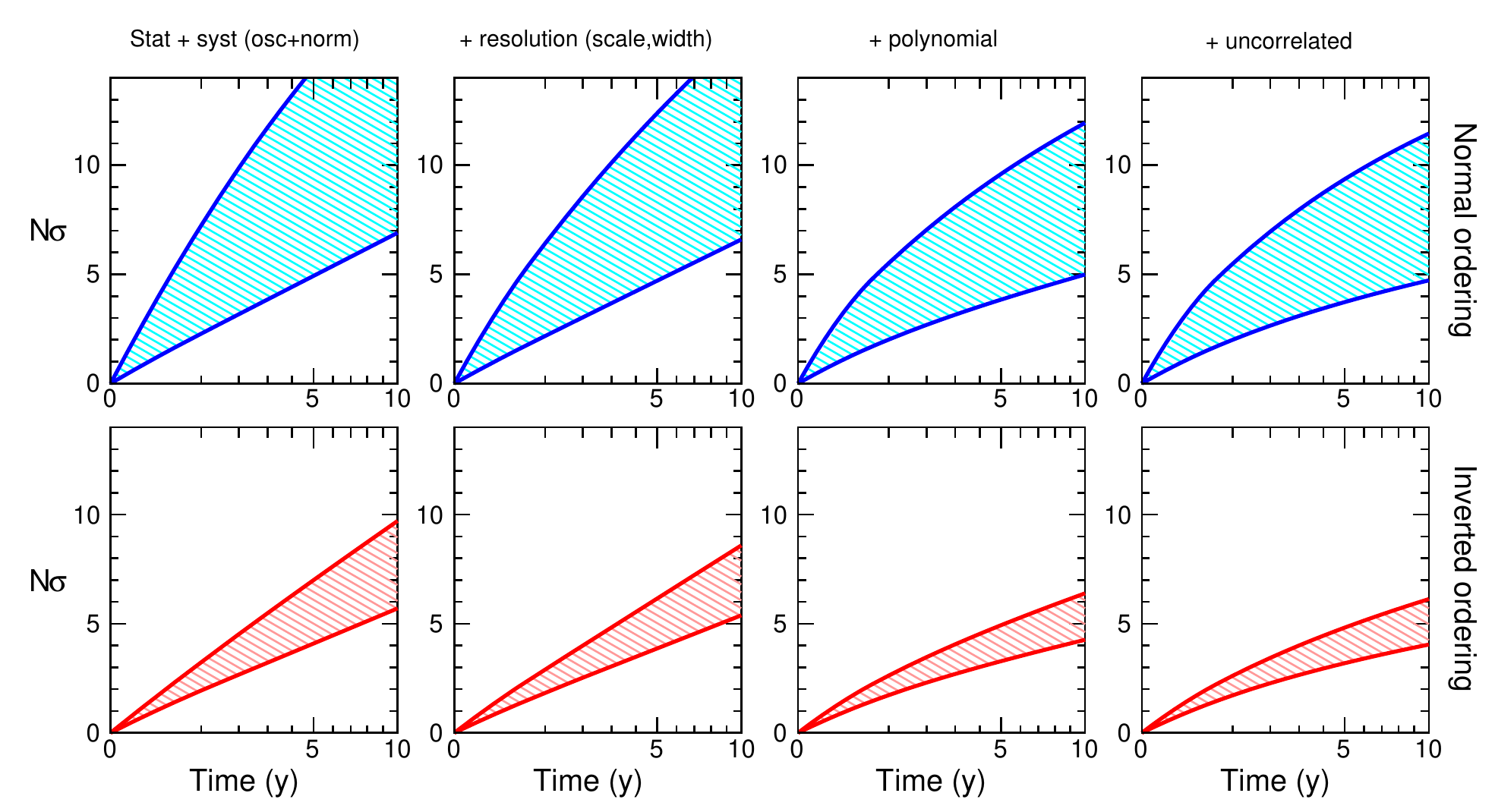}
\vspace*{-7mm}
\caption{As in Fig.~\protect\ref{Fig_03}, but for the hypothetical case of perfect neutrino flavor identification. See the text for details.
\label{Fig_07}}
\end{figure}

It is thus worth considering for ORCA the hypothetical case of perfect neutrino flavor identification,
in which cascade and track event samples contain only CC $e$-like and $\mu$-like events, respectively.%
\footnote{Perfect flavor reconstruction was assumed for simplicity in our previous analysis of PINGU \protect\cite{Build}.}
 Figure~\ref{Fig_07} is analogous to Fig.~\ref{Fig_03}, but for perfect flavor identification. The overall increase of the ORCA
sensitivity to the mass ordering is substantial, and it can be seen at a glance. Although it is unrealistic to think that this hypothetical case 
can be fully realized, the results of Fig.~\ref{Fig_07} show the importance to pursue 
experimental strategies that may   
improve the ``purity'' of cascade and track samples  
(as proxies for electron and muon CC events, respectively).

Numerical results from Figs.~3, 4, 5 and 7 are discussed below.
Table~\ref{table}
shows a synopsis of the ORCA sensitivity to mass ordering, expressed in terms of $N_\sigma$ range
for variable true values of $\sin^2\theta_{23}$, assuming 5 years of data taking. The default case correspond to 
the hypotheses in Fig.~3, namely, polynomial and uncorrelated uncertainties at the level of $1.5\%$, and  
$\sin^2\theta_{23}\in [0.4,\,0.6]$. Variants that can lead to improvements of the sensitivity include:
a reduction of polynomial and uncorrelated systematics by a factor of two; a restriction of the true range 
of $\sin^2\theta_{23}$  in $[0.46,\,0.54]$; an ideal flavor identification (cascade event = electron flavor, 
track event = muon flavor). The ORCA flagship goal of a minimum sensitivity of $3\sigma$ in about 4~years  \cite{Coelho,Heij} 
requires thus to deal with various systematic issues: reduction and better control of spectral shape uncertainties; restriction of the
$\theta_{23}$ range by other oscillation experiments; and last but definitely not least, 
improved event flavor identification. From the viewpoint of statistical errors,
an increase of the effective volume \cite{Kouchner} will also help to reach the goal.

Finally, we briefly compare our results with the mass-ordering sensitivity estimated in \cite{ORCA}.
The analysis in \cite{ORCA} included essentially the uncertainties related to oscillation parameters and to (relative) normalization factors, plus an energy scale uncertainty, for a total of about a dozen nuisance parameters. We have adopted a significantly larger set of systematics, including energy-angle resolution
uncertainties, polynomial spectral shape (correlated) errors, and possible additional uncorrelated errors. As a consequence, our sensitivity estimates are 
generally more conservative than those in \cite{ORCA}. In particular, we find that spectral shape uncertainties tend to reduce the sensitivity by roughly half
standard deviation after five years of data taking (see Table~\ref{table}). Although this may seem a small systematic effect, 
it might require significant extra exposure to be compensated (see Fig.~\ref{Fig_03}), ad thus deserves further investigations. 
Mitigation of spectral errors may be achieved through
an improved understanding of atmospheric neutrino fluxes, cross sections and detector features, as a function of both energy and direction.

\begin{table}[b]
\caption{\label{table}\scriptsize\rm 
Synopsis of the ORCA sensitivity to mass ordering (expressed in terms of $N_\sigma$ range for variable $\sin^2\theta_{23}$)  
with the progressive inclusion of various systematics, for a reference exposure of 5 years in all cases.
The first two numerical columns correspond to the default case, with polynomial and uncorrelated uncertainties at the level of $1.5\%$ and  
$\sin^2\theta_{23}\in [0.4,\,0.6]$ (see Fig.~3).  The other couples of numerical columns correspond to the following variants: halved polynomial and uncorrelated systematics (see Fig.~4, right panels); $\sin^2\theta_{23}\in [0.46,\,0.54]$ (see Fig.~5); perfect
event flavor identification (see Fig.~7).}
\scriptsize\rm 
\centering
\begin{tabular}{l|cc|cc|cc|cc}
\br
 & \multicolumn{2}{c|}{Default case} 
 & \multicolumn{2}{c|}{Halved poly.+uncorr.}
 & \multicolumn{2}{c|}{Reduced $\theta_{23}$ range}
 & \multicolumn{2}{c}{Perfect flavor ident.}\\[1mm]
Errors included in the fit & True NO & True IO & True NO & True IO & True NO & True IO & True NO & True IO \\
\mr
Stat.~+~syst.~(osc.~+~norm.)	&	3.2--7.1 & 2.9--4.2 & 3.2--7.1 & 2.9--4.2 & 3.5--5.8 & 3.5--4.3 & 4.9--14.4 & 4.1--7.0 \\
+ resolution (scale, width)		&	3.0--6.2 & 2.7--3.9 & 3.0--6.2 & 2.7--3.9 & 3.3--5.2 & 3.3--3.9 & 4.7--12.4 & 3.9--6.2 \\
+ polynomial (quartic)			&	2.5--4.6 & 2.3--3.1 & 2.7--5.3 & 2.5--3.4 & 2.6--3.9 & 2.7--3.1 & 3.8--9.6  & 3.3--4.9 \\
+ uncorrelated 
systematics	\ \ \ \ 			&	2.4--4.5 & 2.3--3.0 & 2.7--5.2 & 2.5--3.4 & 2.5--3.7 & 2.6--3.0 & 3.7--9.4  & 3.2--4.8 \\
\br
\end{tabular}
\end{table}

\section{Conclusions and prospects}

Within the KM3NeT collaboration, the large-volume seawater Cherenkov detector ORCA is being proposed, 
in order to observe high-statistics samples of atmospheric neutrino events, that might distinguish the 
normal and inverted neutrino mass orderings through a statistical analysis of energy-angle spectral
variations \cite{ORCA}. Building upon a previous work focused on PINGU \cite{Build}, we have performed herein a thorough 
analysis of prospective ORCA data in the benchmark configuration with 9~m vertical spacing \cite{ORCA}. 
We have discussed the effects of various sources of systematic uncertainties, 
by progressively enlarging the set of nuisance parameters added to statistical errors. Starting
from the most obvious ingredients (oscillation parameter and normalization errors),
we have added errors related to the resolution functions (energy bias plus energy-angle width uncertainties), and
relative spectral shape variations in terms of polynomials (up to quartic dependence in energy and angle), which
are meant to represent the effects of both known and poorly-known sources of spectral shape deformations.
We have also added, on top of such correlated systematics, possible bin-to-bin uncorrelated errors. 
Variations with respect to default choices have been considered, by halving or doubling the size of
some systematics, by changing the prior range of the $\theta_{23}$ mixing angle, and eventually by 
assuming perfect flavor identification. It turns out that, in order to fully exploit the high statistics of cascade and track events
observable in ORCA, it is desirable to control spectral shape systematics at the percent level. 
Improvements in the purity of the event samples (i.e., of cascades and tracks as proxies of electron and muon neutrinos)
will be very beneficial to the mass-ordering sensitivity. 
Better external constraints on $\theta_{23}$ from other experiments will also help, since
ORCA's own measurement of $\theta_{23}$ appears to be entangled with the mass-ordering determination. 
We hope that these finding may trigger further investigations and characterizations of (correlated and uncorrelated, well- and poorly known) 
systematics uncertainties, at the unprecedented level of control required
by future searches of subleading flavor oscillation effects, not only in ORCA but in all very-large-volume atmospheric neutrino experiments.

\ack
The work of F.C.\ is supported by NSF Grant PHY-1404311 to J.F.~Beacom.
The work of E.L.\ and A.M.\  is supported by the Italian 
Ministero dell'Istruzione, Universit\`a e Ricerca (MIUR) and Istituto Nazionale di Fisica
Nucleare (INFN) through the ``Theoretical Astroparticle Physics'' (TAsP) project.

\newpage

\end{document}